\documentclass[%
 aip,
 pop,
 amsmath,amssymb,floatfix,
 reprint,%
]{revtex4-1}

\usepackage{graphicx}% Include figure files
\usepackage{dcolumn}% Align table columns on decimal point
\usepackage{bm}% bold math
\usepackage[utf8]{inputenc}
\usepackage[T1]{fontenc}
\usepackage{mathptmx}
\usepackage{color}

\usepackage{url}

\begin{document}

%\preprint{AIP/123-QED}

\title{Characterizing filamentary magnetic structures in counter-streaming plasmas by Fourier analysis of proton images}

\author{Joseph Levesque}
\email{jmlevesq@umich.edu}
\affiliation{University of Michigan, Ann Arbor, Michigan 48109, USA}
\affiliation{SLAC National Accelerator Laboratory, Menlo Park, CA 94025, USA}

\author{Carolyn Kuranz}
\affiliation{University of Michigan, Ann Arbor, Michigan 48109, USA}

\author{Timothy Handy}
\affiliation{University of Michigan, Ann Arbor, Michigan 48109, USA}
%\affiliation{Radiant Solutions, -}

\author{Mario Manuel}
\affiliation{University of Michigan, Ann Arbor, Michigan 48109, USA}
\affiliation{General Atomics, San Diego, California 92121, USA}

\author{Frederico Fiuza}
\email{fiuza@slac.stanford.edu}
\affiliation{SLAC National Accelerator Laboratory, Menlo Park, CA 94025, USA}

\date{}
	
\begin{abstract}
	Proton imaging is a powerful tool for probing electromagnetic fields in a plasma, providing a path-integrated map of the field topology.
	However, in cases where the field structure is highly inhomogeneous, inferring spatial properties of the underlying field from proton images can be difficult. 
	This problem is exemplified by recent experiments which used proton imaging to probe the filamentary magnetic field structures produced by the Weibel instability in collisionless counter-streaming plasmas. 
	In this paper, we perform analytical and numerical analysis of proton images of systems containing many magnetic filaments. 
	We find that, in general, the features observed on proton images do not directly correspond to the spacing between magnetic filaments (the magnetic wavelength) as has previously been assumed, and that they instead correspond to the filament size.
	We demonstrate this result by Fourier analysis of synthetic proton images for many randomized configurations of magnetic filaments. 
	Our results help guide the interpretation of experimental proton images of filamentary magnetic structures in plasmas.
\end{abstract}

\maketitle
	
\section{Introduction}

\begin{figure*}
	\centering
	\includegraphics{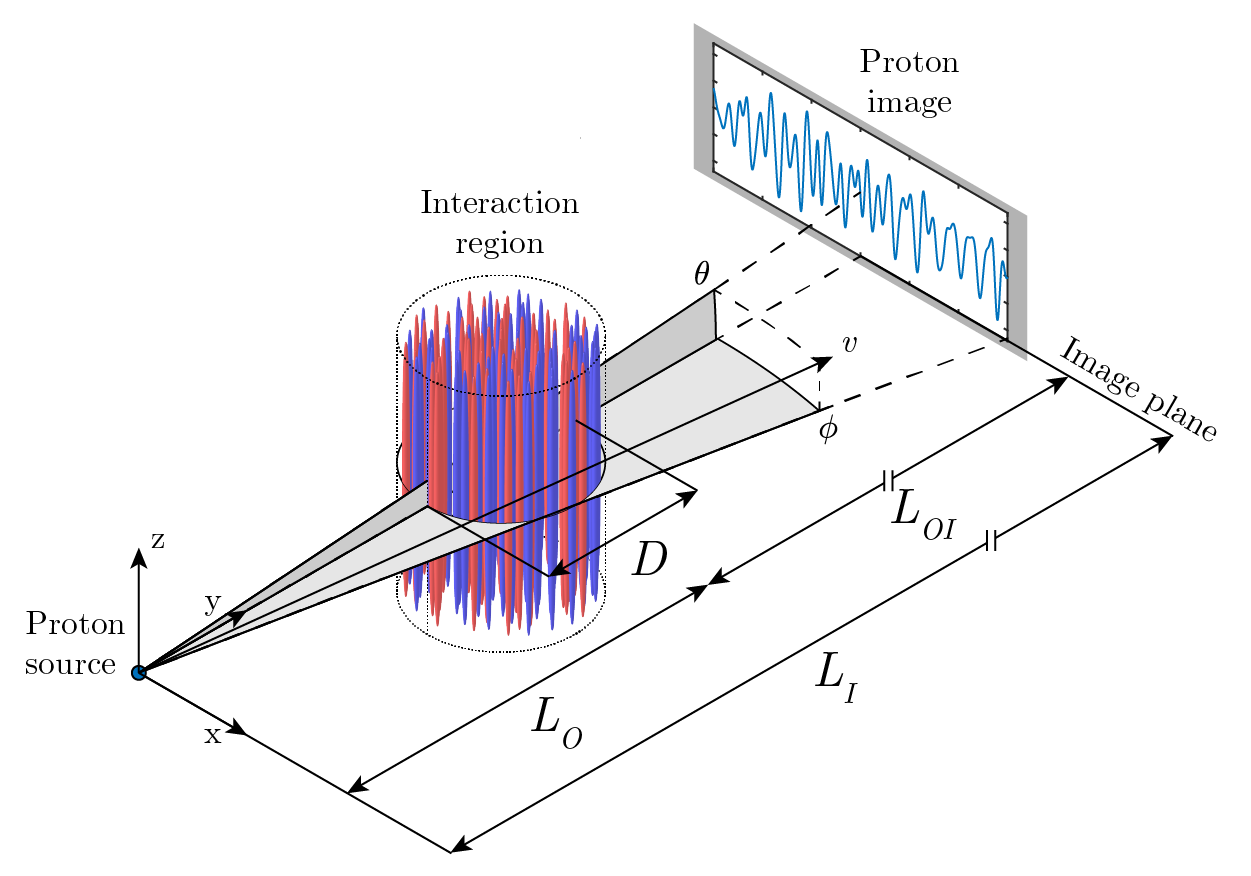}
	\caption{
		Illustration of the system geometry (not to scale).
		The interaction region is cylindrical with radius in the $ xy $ plane and extends infinitely in the $ z $ direction.
		The filaments are contained within the prescribed interaction region, depicted by colored contours of vector potential $ A_z $ for a representative system of filaments.
		Protons stream from the origin with initial velocity $v$ through the interaction region to the image plane.
	}
	\label{fig:SysGeom}
\end{figure*}

Proton imaging is a powerful technique for probing electromagnetic fields in high-energy-density plasma experiments with high-energy ($>$ MeV) protons\cite{Borghesi_PoP2002,Mackinnon_RSI2004b,Li_RSI2006,Li_PoP2009}.
Deflections from initial proton trajectories by interaction with electromagnetic fields encode information about the path-integrated field structure onto a detector in the form of spatial variations of the observed proton flux.
Proton imaging has been used successfully in experiments studying laser-produced plasma bubbles\cite{Borghesi_PoP2002, Li_RSI2006, Li_PoP2009}, magnetic reconnection\cite{Li_PoP2009}, turbulent dynamo amplification of magnetic fields\cite{Tzeferacos_NatCom2018}, and the Weibel instability\cite{Park_HEDP2012,Ross_PoP2012,Fox_PRL2013,Kugland_PoP2013,Huntington_NP2015,Huntington_PoP2017,Ross_PRL2017}, among others.
The detailed analysis by \citet{Kugland_RSI2012} provides a basis for determining proton image structures from electric and magnetic fields, but inferring quantitative field information from an image is difficult due to its path-integrated nature.
Recently, methods have been developed which can infer path-integrated field topology from proton images by solving the inverse problem\cite{Graziani_RSI2017,Bott_JPP2017}.
However, inferring internal structure or spatial scales of magnetic fields which are neither smooth nor homogeneous requires further geometrical assumptions about the system.

An important example of inhomogeneous electromagnetic fields produced in plasmas is the filamentary magnetic field structures associated with the Weibel or current-filamentation instability\cite{Weibel_PRL1959,Fried_PhysFluids1959}.
The Weibel instability is associated with anisotropy of the plasma velocity distribution and is known to lead to the formation of current filaments in counterstreaming, collisionless plasmas, converting kinetic energy into magnetic energy\cite{Weibel_PRL1959}.
This instability is expected to be common in astrophysical plasmas, potentially mediating the amplification of magnetic fields, the formation of collisionless shocks, and the acceleration of particles in energetic and weakly magnetized environments such as gamma ray bursts and young supernova remnant shocks\cite{Medvedev_ApJ1999}.
Furthermore, magnetic fields observed throughout the intergalactic medium may have been seeded by this instability during the early universe\cite{Schlickeiser_ApJ2003,Medvedev_ApJ2006}. 
In recent years, there has been a significant effort to study the Weibel instability and collisionless shocks mediated by it in laboratory laser-driven plasmas\cite{Fox_PRL2013,Huntington_NP2015,Ruyer_PRL2018}. 
Experiments at OMEGA\cite{Park_HEDP2012,Ross_PoP2012,Kugland_PoP2013,Huntington_NP2015,Huntington_PoP2017} and the NIF\cite{Ross_PRL2017} have explored this instability, successfully observing filamentary magnetic fields.

The Weibel instability is an interesting system of study for proton imaging capabilities because the magnetic fields it generates are highly structured, consisting of many small-scale filaments.
The proton images produced when probing these systems perpendicular to the interpenetration axis show filamentary striations in the proton fluence\cite{Kugland_PoP2013,Huntington_NP2015,Park_PoP2015}.
However, inferring internal field parameters from these images has proven to be difficult, because the protons experience deflections from many filaments along any path.
Previous papers\cite{Fox_PRL2013,Park_PoP2015,Huntington_NP2015} infer the characteristic spatial mode of the magnetic field as the average distance between successive peaks of proton fluence on a proton image, but it is unclear if this is an accurate or robust method for characterizing the structure of these fields.
\citet{Levy_RSI2015} explore this issue by creating and analyzing synthetic proton images of randomized distributions of Gaussian magnetic filaments.
The synthetic proton images qualitatively recreate filamentary structures seen in experiment, however, no quantitative relation is determined between the spatial modes of synthetic proton images and those of the corresponding magnetic field.

In this paper we develop a simple analytical model of proton images in the linear deflection regime\cite{Kugland_RSI2012,Bott_JPP2017}, and demonstrate that spatial information of Weibel-like magnetic fields may be inferred from Fourier analysis of proton image features.
This paper is organized as follows. 
Section \ref{sec:ProtonImaging} establishes the methodology which we use to produce synthetic proton images in the linear deflection regime. 
In Section \ref{sec:ForestEffect} we discuss what we call the forest effect, which is associated with probing a large number of filamentary structures, and address the limitations associated with inferring the spacing between filaments by counting the number of peaks in the proton images. 
In Section \ref{sec:FourierAnalysis} we show that the size of filamentary magnetic fields can be inferred from Fourier analysis of the proton images, illustrate that this method is robust for different field configurations, and address the broader applicability of the method we develop.
Section \ref{sec:Conclusion} concludes the paper.

\section{Proton Imaging of Magnetic Filaments} \label{sec:ProtonImaging}

To explore the relationship between filamentary field structures and the associated proton images we develop a simple analytical and numerical model. 
The system geometry consists of three components, the proton source, the interaction region, and the image plane, as illustrated in Figure \ref{fig:SysGeom}.
Following the work of \citet{Kugland_RSI2012} and \citet{Levy_RSI2015}, the filaments are assumed to be Gaussian ellipsoids of vector potential 
\begin{align}\label{eq:Azprime}
\begin{split}
&A_z\left( x',y',z' \right) = A_0 \exp\left( -\frac{x'^2}{a^2}-\frac{y'^2}{a^2}-\frac{z'^2}{b^2} \right), \\
&A_x = A_y = 0,
\end{split}
\end{align}
where $ A_0 $ is the maximum vector potential of the filament, $ a $ is the characteristic filament size in the radial plane, and $ b $ the characteristic length along the collision axis.
The primed coordinates are defined with respect to the center of the filament $(x_c, y_c, z_c)$,
\begin{equation}\label{eq:xprime}
	x' = x-x_c,\quad y' = y - y_c,\quad z' = z - z_c.
\end{equation}
As in \citet{Levy_RSI2015}, all filaments are constrained to lie within the cylindrical volume of the interaction region.
For further simplification, all filaments are oriented along $ \hat{z} $ and considered infinite in length (i.e. $ b \rightarrow \infty $) to remove variation in $z$.

We use the field definition of equation (\ref{eq:Azprime}) to develop a simple analytic model for proton images of filamentary magnetic fields in the linear proton deflection regime.
The linear deflection regime, as defined by \citet{Kugland_RSI2012}, is when proton deflections are small relative to the scale length of electromagnetic fields.
The linearity parameter (adapted for our notation) is defined as 
\begin{equation}\label{eq:linearity}
	\mu \equiv L_{O} \alpha / L_{EM},
\end{equation}
where $L_{O}$ is the $ y $-distance from the proton source to the center of the interaction region, $\alpha$ is the proton deflection angle, and $L_{EM}$ is the scale length of the electromagnetic fields. 
The linear regime is defined as when $\mu \ll 1$.
In the case of filamentary magnetic fields, $L_{EM} = a$, and the linear regime applies when $\alpha \ll a/L_{O}$.
In the linear regime, deflections from a proton's initial, unperturbed trajectory are negligible across a single filament, so the total deflection can be calculated from the path-integrated vector potential along the unperturbed trajectory.

We consider a point-like, divergent proton source with initial proton velocities
\begin{equation}\label{eq:velocity}
	v_x = v \cos\theta \sin\phi, \quad
	v_y = v \cos\theta \cos\phi, \quad
	v_z = v \sin\theta,
\end{equation}
where $v$ is the initial speed and the angles $\theta$ and $\phi$ are angles from the $y$ axis in the $yz$ and $xy$ plane as depicted in Figure 1.
The angle $\phi$ provides for magnification effects in the $x$ direction.
Because of the uniformity and infinite extent of the filaments along $z$, we consider only proton deflections in the $x$ direction, producing 1D proton images at a nonzero angle $\theta$ as depicted in Figure \ref{fig:SysGeom}.
Furthermore, we consider the small deflection regime, where the interaction region diameter $D$ is much smaller than the distance $L_O$ between the proton source and the interaction region, which is typically true in experiment.
In this paraxial limit, proton trajectories are considered constant across the interaction region. 
The angular deflections in $x$ are calculated as
\begin{align} \label{eq:alphax}
\begin{split}
	\alpha_x = \frac{e}{c\sqrt{2 m_p W}} \sin\theta \frac{\partial}{\partial x} \int_{0}^{L_I}A_z \text{d}y,
\end{split}	
\end{align}
where $ e $ is the electric charge, $ m_p $ is the proton mass, $ c $ is the speed of light, and $W$ is the energy of the probe protons.
We note that in the limiting case of $\theta = 0$, equation (\ref{eq:alphax}) predicts no deflection in $x$.
In reality, small deflections in $x$ can still arise from higher-order terms as shown in \citet{Kugland_RSI2012}.
However, in typical experiments the proton source is divergent, providing the necessary $v_z$, and equation (\ref{eq:alphax}) can thus be used to describe the dominant deflections.

The entire length $D$ across the interaction region is contained within the integration bounds of equation (\ref{eq:alphax}), and we assume a single filament $a \ll D$, so we can approximate the integral as
\begin{equation}\label{eq:IntAzxy}
\int_{0}^{L_I} A_z \text{d}y \approx \int_{-\infty}^{\infty} A_z \text{d}y = \sqrt{\pi a^2} A_0 \exp \left(  -\frac{x^2}{a^2} \right).
\end{equation}
Substituting into equation (\ref{eq:alphax}), the deflection by a single filament is
\begin{equation}
	\alpha_x\left(x\right) = \frac{e}{c\sqrt{2 m_p W}} A_0 \sin\theta \left( -\frac{2\sqrt{\pi}x}{a} \right) \exp\left( -\frac{x^2}{a^2} \right).
\end{equation}

Assuming the deflection occurs at the center of the interaction region, protons arrive at the image plane with deflected positions
\begin{equation}\label{xImage}
x_{I}\left( x \right) = \frac{L_{OI}}{L_{O}}\left[ x + L_{O}\alpha_x \left(x\right)\right],
\end{equation}
where $ x $ is the position of a proton along an unperturbed trajectory defined at $ y=L_{O} $.
From \citet{Kugland_RSI2012}, in the linear deflection regime the proton fluence map $ I $ at the image can be calculated as
\begin{equation}\label{eq:ILinear}
	I = I_0\left[ 1-L_{O}\frac{\partial}{\partial x}\alpha_x \right],
\end{equation}
where $ I_0 $ is the initial, unperturbed proton fluence profile.
The proton image from deflections by a Gaussian filament is thus
\begin{equation}\label{eq:Image}
I\left( x \right) = I_0 - L_O \frac{e}{c} \sqrt{\frac{\pi a^2}{2 m_p W}} A_0 \sin\theta \frac{\partial^2}{\partial {x}^2} \exp\left( -\frac{x^2}{a^2}\right),
\end{equation}
whose profile is primarily proportional to the second derivative of the path-integrated vector potential $A_z$ --- the second derivative of a Gaussian in $x$.
By normalizing about the mean image intensity, the influence of any Gaussian filament on the image can be generalized to
\begin{equation}\label{eq:dI2}
	\delta I \left( x,x_c \right) = \left( -\frac{2}{a^2}+\frac{4 \left({x-x_c}\right)^2}{a^4} \right)e^{-\left({x-x_c}\right)^2/a^2}.
\end{equation}

In the linear deflection regime the image contribution from each filament is effectively independent from one another.
Thus, the image of a system of multiple filaments is the summation of contributions from all filaments in the system
\begin{equation}\label{eq:ImageSum}
I\left( x \right) = I_0\left( x \right) + \sum_{n}^{N} \delta I\left(x,x_{cn}\right) = I_0\left( x \right) + \sum_{n}^{N} \delta I_n\left( x \right) ,
\end{equation}
where the subscript $ cn $ refers to the center of a filament $n$.

\section{The Forest Effect} \label{sec:ForestEffect}

\begin{figure}
	\centering
	\includegraphics{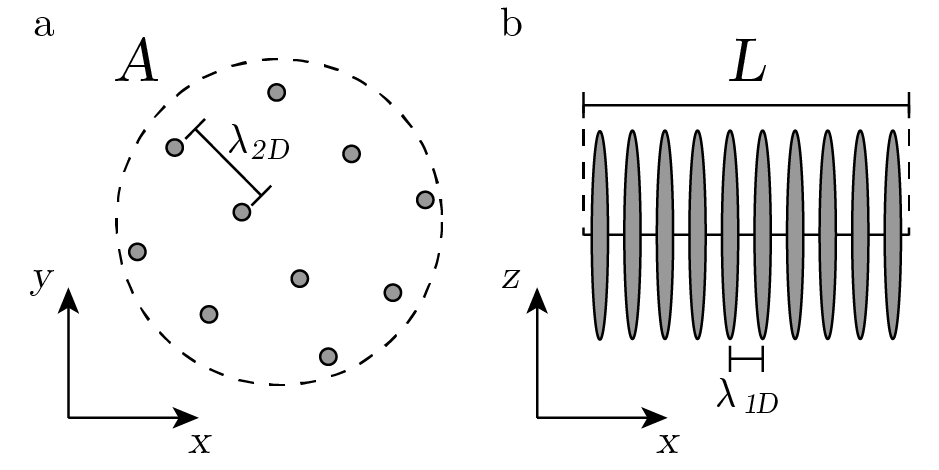}
	\caption{
		Spatial scales for (a) top-down view showing an areal distribution of filaments, and (b) a side-on view showing a linear distribution of the same number of filaments.
	}
	\label{fig:StatSpacing}
\end{figure}

%Visual guide of determining proton images computationally
\begin{figure*}
	\centering
	\includegraphics{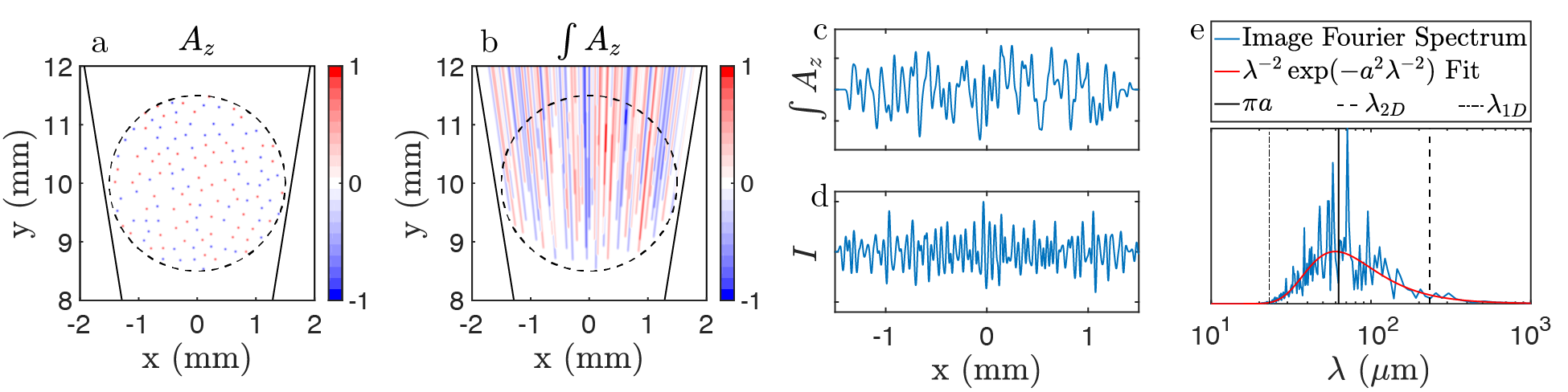}
	\includegraphics{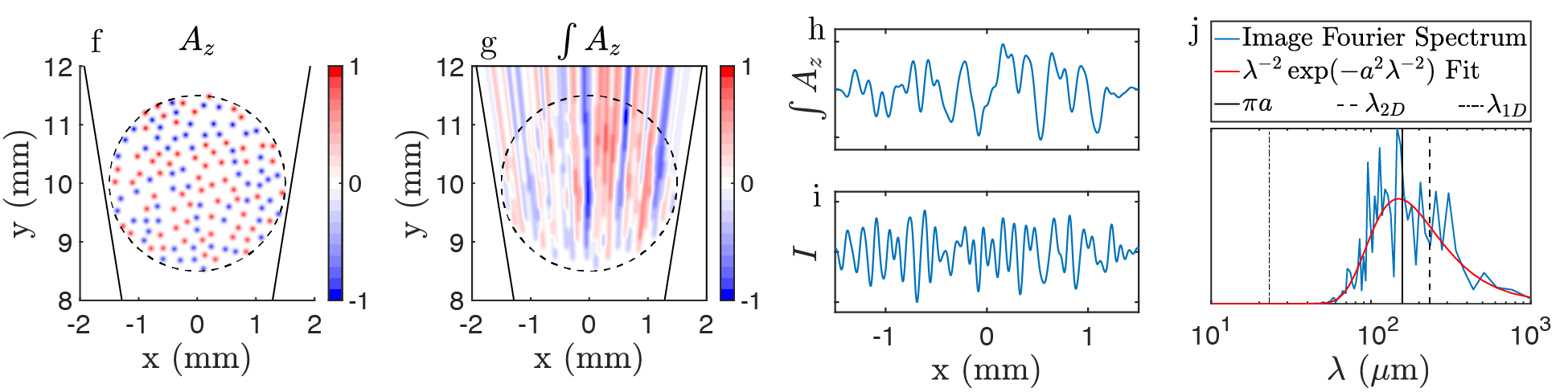}
	\caption{
		Representative results of (a) the normalized vector potential map, (b) the normalized path-integrated vector potential map, (c) the path-integrated vector potential plot at image, (d) the resulting proton fluence at an angle $ \theta =0.1$, and (e) the $ \lambda $-space Fourier spectrum for a synthetic distribution of 130 filaments with $ a= 20 $ $ \mu $m.
		Repeated for $ a= 50 $ $ \mu $m in (f-j) for the same centroid distribution.
		The analytic fit of the synthetic Fourier spectra (red) tracks with filament size $ a $ following the derived analytic relation.
	}
	\label{fig:DistributionsDoNotMatter}
\end{figure*}

\begin{figure}
	\centering
	\includegraphics{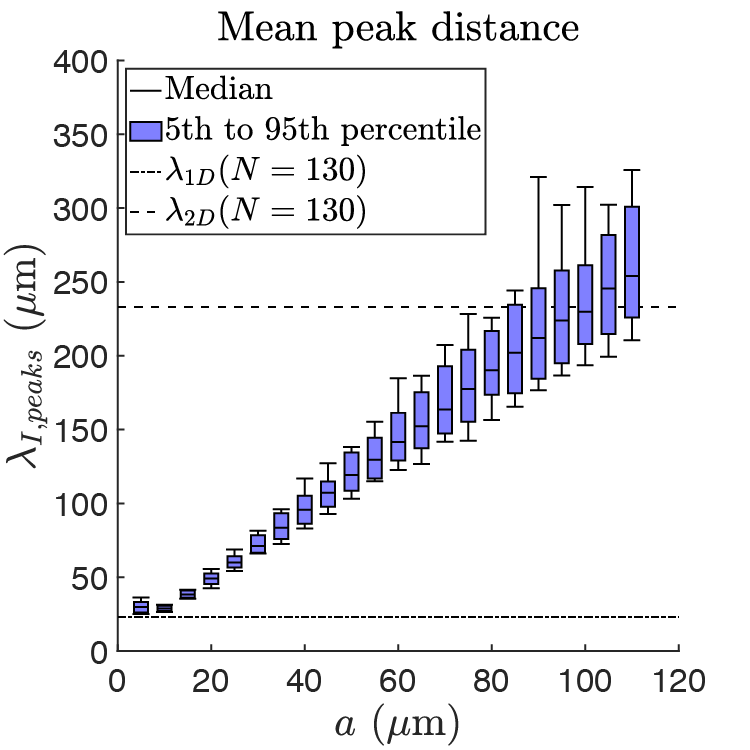}
	\caption{
		Results of the mean peak distance analysis method for 100 randomized distributions of 130 identical filaments at each prescribed filament size $a$.
		The vertical bars represent the range of inferred values of $\lambda_{I,peaks}$, and the boxes represent the range in which 90\% of the inferred values lie.
		The lines of $\lambda_{1D}$ and $\lambda_{2D}$ are obtained by equations (\ref{lambda1D}) and (\ref{lambda2D}) for $N=130$ and $D=3$ mm.
	}
	\label{fig:PeakSpacing}
\end{figure}

Previous works have assumed that the spacing between filaments in the interaction region can be directly inferred from the average spacing between successive peaks in fluence on a proton image, adjusted for magnification\cite{Fox_PRL2013,Huntington_NP2015}.
This may seem like a simple and attractive solution, but because of the complexity of the underlying field structures it is not clear whether this method accurately infers the spacing between filaments in the plasma as intended.

Consider the problem more simply: envision a forest of trees whose centroids are described by a spatial Poisson point process.
When viewed from above (Figure \ref{fig:StatSpacing}a) the spatial modes of this system may be characterized by the average distance between centroids in the $ xy $ plane, defined as
\begin{equation}\label{lambda2D}
\lambda_{2D} = \sqrt{\frac{A}{N}},
\end{equation}
where $A$ is the 2D area in which $ N $ trees exist.
When looking instead at this forest from the ground along the $ y $ direction (Figure \ref{fig:StatSpacing}b), position information collapses along that axis.
Ignoring magnification effects, the position of the trees can now only be discerned in one dimension --- along $ x $.
The spatial mode in this case corresponds to the average $ x $ distance between the trees
\begin{equation}\label{lambda1D}
\lambda_{1D} = L/N,
\end{equation}
where $ L $ is the length across the forest.
Now extend this line of thought to a distribution (a \emph{forest}) of identical magnetic filaments, or more specifically, the proton image of such a forest of filaments.
This simple example illustrates that, naively, one would expect that when the number of filaments (trees) is very large ($N \gg 1$), the wavelength of the filamentary magnetic structures (the spacing between trees) that is inferred from the proton image is significantly smaller, by a factor of $\sqrt{N}$, than the actual filament wavelength. 

To explore whether the method of counting peaks on proton images accurately infers the spacing between filaments, we generate randomized distributions of filaments from which we create synthetic 1D proton images.
The synthetic system geometry corresponds to the OMEGA experiments\cite{Fox_PRL2013,Huntington_NP2015,Levy_RSI2015}; the interaction region is an infinite cylinder of diameter $D=3$ mm, $ L_O=1 $ cm, $ L_{OI}=30 $ cm, and the image plane is a $ 9.6 $ cm x $ 9.6 $ cm square.
The synthetic proton images are created for $\theta = 0.1$, and have a resolution of 45 $\mu$m at the image plane, corresponding to 1.5 $ \mu $m with respect to the interaction region.
The filament centroids in the interaction region are randomized for each distribution, similar to the setup of \citet{Levy_RSI2015}, with an enforced minimum distance between each filament to prevent overlap.
To conform to expected physical constraints, there are an equal number of positive and negative $ A_z $ filaments in every distribution. 
The filament size $ a $ is varied independently of the system size and number of filaments, though these may be related in reality.
Figures \ref{fig:DistributionsDoNotMatter}a and \ref{fig:DistributionsDoNotMatter}f show the vector potential of a typical distribution of 130 filaments for $ a $ = 20 $ \mu $m and $ a $ = 50 $ \mu $m, respectively.
For each distribution of filaments we numerically integrate the vector potential along diverging proton paths to the image plane, as shown in Figures \ref{fig:DistributionsDoNotMatter}b and \ref{fig:DistributionsDoNotMatter}g, to account for magnification effects.
We use the integrated vector potential to calculate $\alpha_x$, and generate the synthetic 1D proton image (Figures \ref{fig:DistributionsDoNotMatter}d and \ref{fig:DistributionsDoNotMatter}i) using equation (\ref{eq:Image}).
By counting peaks in the proton image, we infer an average spacing between consecutive peaks as
\begin{equation} \label{eq:lambdapeaks}
	\lambda_{I,peaks} = D/N_{I,peaks},
\end{equation}
where $N_{I,peaks}$ is the number of peaks on the image.

%Figures demonstrating the different methods and their accuracy
\begin{figure*}
	\centering
	\includegraphics{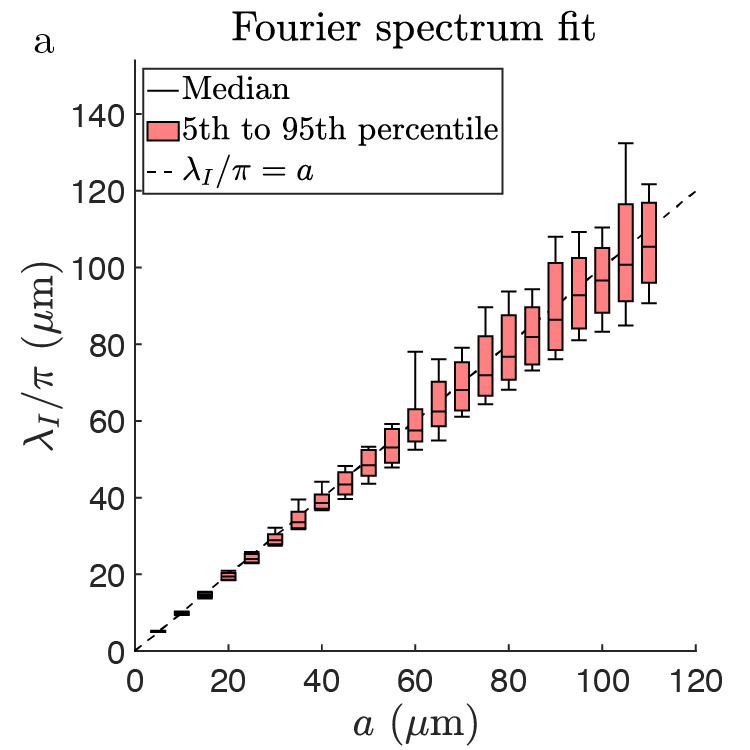}
	\includegraphics{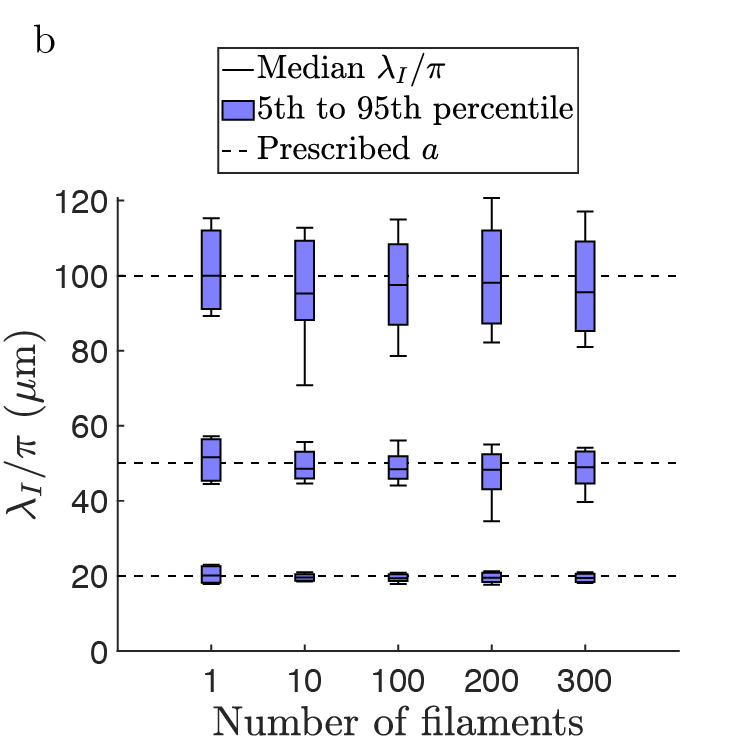}
	\caption{
		(a) Statistics of fitting the image Fourier spectrum to equation (\ref{Fit}) for 100 randomized distributions of 130 identical filaments for many values of filament size $ a $.
		(b) Fourier spectrum analysis statistics for 100 randomized distributions for different numbers of filaments and prescribed $a$ of 20, 50, and 100 $\mu$m.
		Fitting the Fourier spectrum to our analytic relation reliably infers the prescribed filament size from synthetic proton images for all $ a $.
	}
	\label{fig:MethodComp}
\end{figure*}

The results of this analysis are shown in Figure \ref{fig:PeakSpacing} for systems containing 130 filaments.
We analyze 100 randomized distributions for each prescribed value of $a$ to average over any distribution dependence.
The median inferred $\lambda_{I}$ approximately reaches the limit $\lambda_{1D}$ at $\lambda_{1D} \approx 2a$.
Above this limit $\lambda_{I,peaks}$ increases linearly with $a$ and does not appear dependent on $\lambda_{2D}$ even at large $a$.
We therefore conclude that the forest effect is only present for cases when $a\leq \lambda_{1D}/2$, for which $\lambda_{I,peaks}\approx \lambda_{1D}$.
For the case of the Weibel instability, the magnetic wavelength can be approximated as $\lambda_W \sim 4a$\cite{Ruyer_PRL2018}, and thus $2a/\lambda_{1D}=\sqrt{\pi N}/2$.
For most cases of interest $N \gg 1$, which means $2a/\lambda_{1D} > 1$, and thus $\lambda_I$ will not correspond to $\lambda_{1D}$.

\section{Inferring Filament Size via Fourier Analysis} \label{sec:FourierAnalysis}

\subsection{Analytic Solution} \label{sec:AnalyticSolution}

In this section we develop a simple method to determine the spatial size of the magnetic filaments from Fourier analysis of the proton images.
We return to the simple analytic model of Section \ref{sec:ProtonImaging} and derive the expected dominant spatial mode of the proton image of a system containing many filaments.
By Fourier transforming the proton image defined in equation (\ref{eq:ImageSum}) we find
\begin{equation}\label{FfN}
\mathcal{F}\left\lbrace I \right\rbrace = \sum_{n}^{N} \sqrt{\frac{a^2}{2}} k^{2} \exp\left( -\frac{a^2}{4}k^2 + ix_{cn}k \right).
\end{equation}
The phase terms $i x_{cn} k$ in equation (16) carry information about filament position, and add deviations to the spectrum.
However, when summing over the contributions from many randomized filaments these deviations will be small relative to the overall Fourier profile
\begin{equation}\label{Ff2}
\mathcal{F}\left\lbrace I \right\rbrace = C k^{2} \exp\left( -\frac{a^2}{4}k^2 \right),
\end{equation}
which will be present for any distribution.
We use this phase-free profile to find the dominant image mode
\begin{equation}\label{kzero}
k_I = 2 / a,
\end{equation}
from which the dominant observed image wavelength is
\begin{equation}\label{lambdaI}
\lambda_{I} = 2\pi / k_{I} = \pi a.
\end{equation}
Thus, $ \lambda_I $ for a system of identical filaments in the linear deflection regime will primarily depend on the size of the constituent filaments and not the separation between them.

\subsection{Statistical Verification} \label{sec:StatisticalVerification}

To test the analytic relation of equation (\ref{lambdaI}), we again generate randomized distributions of filaments and analyze corresponding synthetic 1D proton images.
Now, however, we use equation (\ref{Ff2}) to determine $ \lambda_I $ by fitting the function 
\begin{equation}\label{Fit}
\mathcal{F}\left\lbrace I \right\rbrace = C\lambda^{-2} e^{-a^2 /\lambda^{2}},
\end{equation}
to the Fourier spectrum of the image.
The maximum value of this fit occurs at $ \lambda_{I} $, from which we infer the filament size as 
\begin{equation}\label{lambdarelation}
a_{I} = \lambda_{I} / \pi.
\end{equation}
The Fourier spectra in Figures \ref{fig:DistributionsDoNotMatter}e and \ref{fig:DistributionsDoNotMatter}j show large modulations, but the fit of equation (\ref{Fit}) accurately determines $\lambda_I$ very close to $ \pi a $ for the same centroid distribution.

%Figures of the GRF images for 20 and 50 micron
\begin{figure*}
	\centering
	\includegraphics{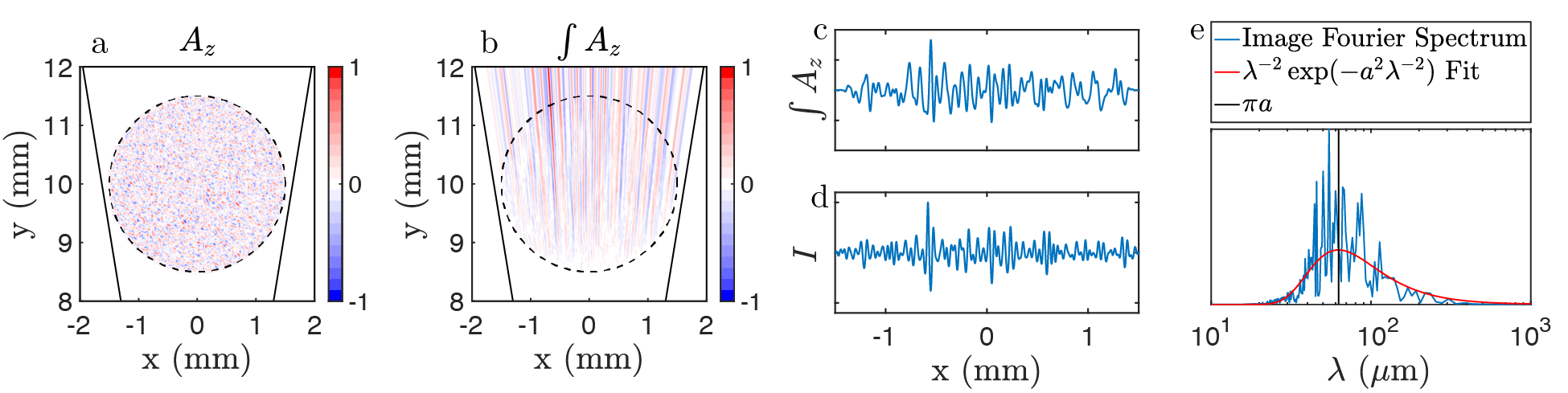}
	\includegraphics{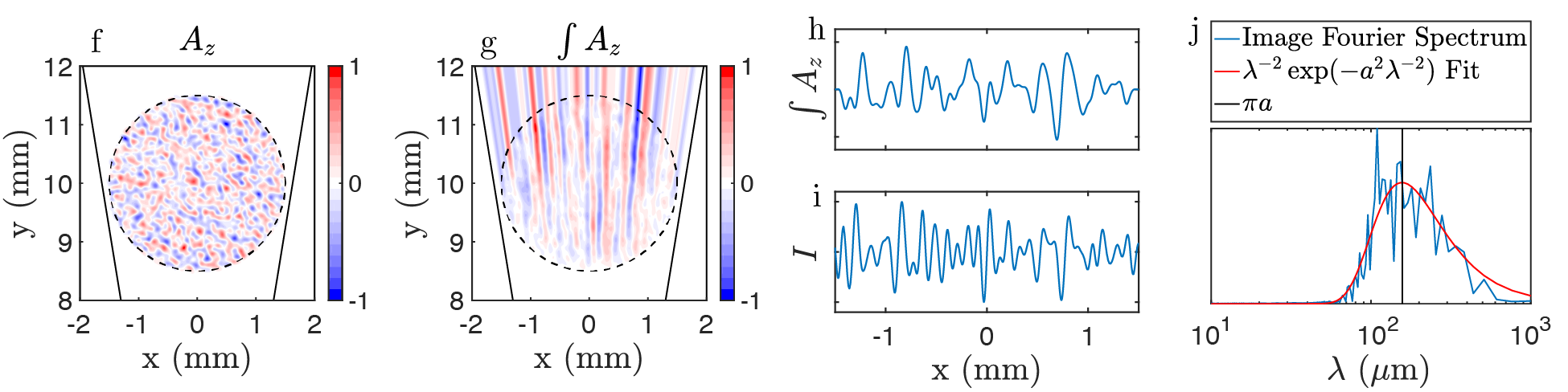}
	\caption{
		Representative results of (a) the normalized vector potential map, (b) the normalized path-integrated vector potential map, (c) the path-integrated vector potential plot at image, (d) the resulting proton fluence at an angle $ \theta=0.1 $, and (e) the $ \lambda $-space Fourier spectrum for Gaussian random fields generated with $ a= 20 $ $ \mu $m.
		Repeated for $ a= 50 $ $ \mu $m in (f-j).
		The fit of the Fourier spectra again depends only on the prescribed filament size.
	}
	\label{fig:GRFSeries}
\end{figure*}

We test the accuracy of this method by analyzing many randomized distributions of filaments, as in Section \ref{sec:ForestEffect}.
Figure \ref{fig:MethodComp}a displays the results of this Fourier analysis method for 100 randomized distributions of 130 identical filaments for a range of independently varying $ a $. 
The median inferred $ a_I $ for these distributions is within $ \sim 5\% $ of the prescribed value of $ a $, with $ 90\% $ of the distributions within $ 15\% $.
We also test whether the number of filaments in the system affects the accuracy.
Figure \ref{fig:MethodComp}b shows the inferred $ a_{I} = \lambda_{I}/\pi $ for 100 distributions of systems with 1 to 300 identical filaments at a fixed $ a $ = 20, 50, and 100 $ \mu $m.
From Figures \ref{fig:MethodComp}a and \ref{fig:MethodComp}b, we find that this method accurately infers the prescribed filament size $ a $, regardless of the number of filaments or filament size.

Within the assumptions made throughout this paper, the results displayed in Figures \ref{fig:DistributionsDoNotMatter} and \ref{fig:MethodComp} conclusively demonstrate that the effects of the independent filaments are cumulative, and filament position information is effectively lost.
Instead, the proton images provide direct information about the individual size of each filament, which can be directly obtained from Fourier analysis of the proton fluence profile. 
We note that in practice, for the case of the Weibel instability, the filament size will be directly related to the spacing between filaments, thus one can infer the scale of the magnetic wavelength by measuring the filament size $a$. 

\subsection{Gaussian Random Fields} \label{sec:GRF}

\begin{figure}
	\centering
	\includegraphics{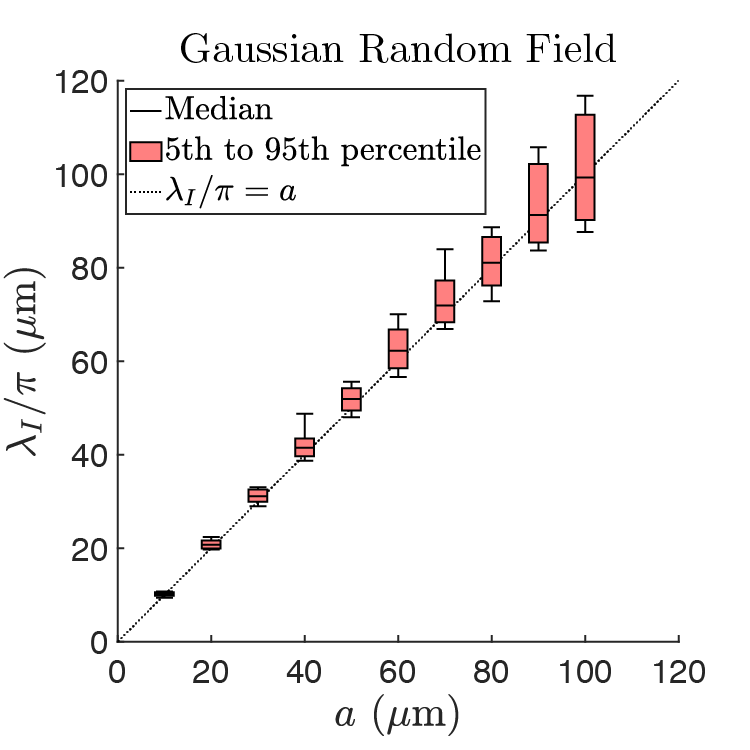}
	\caption{
		Statistics of the Fourier analysis method for 100 initializations of the vector potential as a Gaussian random field at different values of filament size (correlation length) $ a $.
		Fitting the Fourier spectrum to the analytic relation again reliably infers the prescribed filament size from synthetic proton images for all $ a $. 
	}
	\label{fig:GRFStats}
\end{figure}

A more realistic model for the magnetic field profile produced by the Weibel instability is to define the vector potential at the interaction region as a spatial Poisson process, or Gaussian random field.
We follow the method described by \citet{Kroese_LNM2015} and \citet{Dietrich_SIAM1997} to create a stationary, zero-mean, two-dimensional Gaussian process of vector potential using the covariance function
\begin{equation}\label{eq:GRFcov}
\rho\left( x,y \right) = \left[ 1 - \frac{x^2}{l_x^2} - \frac{x y}{l_x l_y} - \frac{y^2}{l_y^2}\right] \exp\left( - \frac{x^2}{l_x^2} - \frac{y^2}{l_y^2} \right),
\end{equation}
where $l_x$ and $l_y$ are the lengths over which the vector potential is correlated in $x$ and $y$, respectively.
To correlate the vector potential by the area of a generalized filament $\pi a^2$ we set $l_x , l_y = \sqrt{\pi} a$ and rewrite equation (\ref{eq:GRFcov}) as
\begin{equation}\label{eq:GRFcov2}
\rho\left( x,y \right) = \left[ 1 - \frac{\left(x+y\right)^2}{\pi a^2}\right] \exp\left( - \frac{x^2}{\pi a^2} - \frac{y^2}{\pi a^2} \right),
\end{equation}
which is reminiscent of how we defined our Gaussian filaments in equation (\ref{eq:Azprime}).
Figures \ref{fig:GRFSeries}a and \ref{fig:GRFSeries}f show representative fields produced by this method at $ a = 20 $ $ \mu $m and $ a = 50 $ $ \mu $m.
We again assume that the protons are negligibly deflected within the interaction region and follow the same process for generating proton images as in Figures \ref{fig:GRFSeries}d and \ref{fig:GRFSeries}i.
The fit of equation (\ref{Fit}) to the Fourier spectra accurately determines the maximum $ \lambda $ to be $ \pi a $, just as before.
Figure \ref{fig:GRFStats} shows that this method reliably infers the prescribed filament size, or spatial correlation length, from the synthetic proton just as it did for systems of individual filaments.

By modeling the magnetic vector potential of filaments created by the Weibel instability as a two-dimensional Gaussian random field, it becomes clear why we can recover the filament size parameter from proton imaging: the field is zero-mean, and integrating across a probing direction (e.g. $ y $) essentially reduces the field to a one-dimensional Gaussian process with covariance function
\begin{equation}\label{GRF1Dcov}
\rho\left( x \right) = \left( 1 - \frac{x^2}{\pi a^2} \right) \exp\left( - \frac{x^2}{\pi a^2} \right),
\end{equation}
which reproduces the Gaussian structures observed.
This analysis suggests that when probing the filamentary fields produced by the Weibel instability, and more generally for any magnetic field produced Gaussian random vector potential, in the linear proton deflection regime, that the structure size parameter can be inferred from the proton image.

%GRF statistics figure

\subsection{Applicability} \label{sec:Applicability}

Our analysis is derived for proton images in the linear proton deflection regime, and it is thus important to clarify the limits of its applicability.
For proton radiography nearly perpendicular to the magnetic filaments, the linearity condition can be written as
\begin{equation}
	B_0(\mathrm{T}) \lesssim 13.4\sqrt{W(\mathrm{MeV})}/L_O(\mathrm{cm}).
\end{equation}
For typical proton radiography parameters, $W=14.7$ MeV and $L_O = 1$ cm, we expect our analysis to be valid for $B_0<50$ T, in agreement with the limit expressed in \citet{Kugland_RSI2012}
We have confirmed this by simulating proton images of filament systems at varying magnetic field strengths.
Above this limit we enter the caustic regime\cite{Kugland_RSI2012}, and the Fourier spectrum starts to be significantly modified.
The possibility of extending this analysis to the caustic regime is outside the scope of this paper, and will be explored in future work.

We also note that while we have only used 1.5 micron resolution in the interaction region for our Fourier analysis, an accurate fit will generally be possible as long as the resolution is smaller than $\pi a/2$ to resolve the Fourier peak.
In practice the image resolution with respect to the interaction region is limited by the proton source size, which can vary from a few microns for protons generated via Target Normal Sheath Acceleration \cite{Hatchett_PoP2000} to $\sim 40$ $\mu$m for protons produced by the implosion of a fusion capsule\cite{Li_RSI2006}. 
This should be carefully considered in application to future experiments, because at typical experimental conditions $a \approx 50$--$100$ $\mu$m\cite{Huntington_NP2015}.

\section{Conclusion} \label{sec:Conclusion}

For systems of independent Gaussian magnetic filaments we have shown that, in the limit of linear proton deflections, proton images primarily provide information about the individual size of the filaments, not the spacing between them.
We have developed a simple analytical model for the linear deflection of protons and have shown that Fourier analysis of the proton images allow for an accurate measurement of the filament size, independent of the number or density of filaments.
Statistical computational analysis of synthetic proton images for many randomized distributions of magnetic filaments shows that this method accurately infers the prescribed filament size.
Additionally, we have shown that when modeling the vector potential as a Gaussian random field, in which the effective filament size and number of filaments are correlated, our analysis produces the same behavior in the limit of linear proton deflection.
This simple method and underlying analysis provides a robust way to characterize proton images of filamentary magnetic fields and should be broadly applicable to proton imaging of magnetic fields in plasma experiments.

\section{Acknowledgements}

F. F. thanks D. D. Ryutov for discussions that motivated this work. 

This work was supported by the U.S. Department of Energy SLAC Contract No. DE-AC02-76SF00515, by the U.S. DOE Office of Science, Fusion Energy Sciences under FWP 100182 and FWP 100237, and by the U.S. DOE Early Career Research Program under FWP 100331.
This work was also supported by the U.S. Department of Energy, through the NNSA-DS and SC-OFES Joint Program in High-Energy-Density Laboratory Plasmas, grant number DE-NA0002956, and the National Laser User Facility Program and William Marsh Rice University, grant number, R19071, and through the Laboratory for Laser Energetics, University of Rochester by the NNSA/OICF under Cooperative Agreement No. DE-NA0001944.
	
\bibliography{References}
\end{document}